# A Relational Triple Extraction Method Based on Feature Reasoning for Technological Patents

Runze Fang, Junping Du*, Yingxia Shao, Zeli Guan

School of Computer Science, Beijing Key Laboratory of Intelligent Telecommunication Software and Multimedia, Beijing University of Posts and Telecommunications, Beijing 100876, China

**Abstract:** The relation triples extraction method based on table filling can address the issues of relation overlap and bias propagation. However, most of them only establish separate table features for each relationship, which ignores the implicit relationship between different entity pairs and different relationship features. Therefore, a feature reasoning relational triple extraction method based on table filling for technological patents is proposed to explore the integration of entity recognition and entity relationship, and to extract entity relationship triples from multi-source scientific and technological patents data. Compared with the previous methods, the method we proposed for relational triple extraction has the following advantages: 1) The table filling method that saves more running space enhances the speed and efficiency of the model. 2) Based on the features of existing token pairs and table relations, reasoning the implicit relationship features, and improve the accuracy of triple extraction. On five benchmark datasets, we evaluated the model we suggested. The result suggest that our model is advanced and effective, and it performed well on most of these datasets.

## 1 Introduction

Entity relation triples extraction is a basic task in information extraction. It attempts to extract triples from unstructured text [1]. Relational Triple Extraction (RTE) is useful for many downstream applications. For example, knowledge reasoning and knowledge graph are very important [2][3].

The conventional pipeline method extracts entities firstly, then categorizes the different types of relationships in predefined set of relationships. However, these models neglect the connection between the two tasks and are vulnerable to bias propagation as a result of the total separation of entity recognition and relationship classification [4].

At present, the main method of RTE is the method of extracting entities and relationships simultaneously, which is called the joint extraction method as well [5][6]. This approach addresses the issue of bias propagation. On several benchmark datasets, some of the most recent joint extraction methods demonstrate remarkable extraction capabilities, particularly when it comes to complicated phrases which contain numerous or overlapping triples.

Among these current methods for joint extraction, a method based on table filling performs exceptionally well. Each item in the table is used to determine if the token pair has a matching relationship in these methods, which normally keep a table for each relationship. As a result, the core to these methods is to precisely complete the relational table, and extract triples based on the full table. However, the existing methods to populate the relationship table are mainly based on extracting features from a single token pair or a single relationship table, but ignore the associations and different relationships between various entity pairs in the same input sentence.

To make full advantage of the relationships between different entity pairs and between different relational tables, we provide a table-filling-based feature reasoning relational triple extraction model. Based on the model, the historical features of token pairs and table relations are integrated. The features can better reveal the differences between relationships and token pairs, which can not only improve the accuracy of triplet extraction through multiple mutual verification, but also improve the recall rate of triplet extraction by helping to derive new triples. For each relationship, we first create a table feature. In a related table feature and a subject object related feature, all relational tables' features are then combined. On this basis, the method based on transformer is used to reasoning entity relationship features.

In summary, the following are the primary contributions of our work:

We suggest an entity relationship extraction method for the technological patent data set.

We proposed a new table filling strategy and triple decoding method.

We integrate previous relationship table features and subject object features to reasoning the implicit relationship features.

## 2 Related Work

Early studies often adopted a pipeline based RTE method [7][8]. This approach initially identifies every entity in the input text before predicting the relationships between every pair of entities. However, pipeline-based methods have two serious drawbacks [9][10]. The correlation between entity recognition and relationship prediction is ignored. They also frequently encounter the issue of error propagation. Researchers started investigating joint extraction methods that concurrently extract entities and connections in attempt to address these drawbacks [11][12].

*Corresponding author: Junping Du (junpingdu@126.com).

One of joint extraction methods is tagging based methods [13] , which usually extracts entities through labels first, and then predicts relationships. These models often employ binary mark sequences to establish the beginning and ending locations of entities, as well as often to establish the connection between two entities [14][15]. By using a position-aware attention method to tag a single text n times, Dai [16] extract triplets successfully. Tan used a ranking with translation approach to accomplish this challenge. Takanobu used reinforcement learning to first identify relations and then recognize entity pairings [17][18][20]. Sequence-to-sequence methods belong to extraction methods as well. This type of approach [17] frequently transforms the relational triple extraction into a task of producing triples in a certain order, such as generating relationships and then generating entities [21][22][23]. To decode overlapping connections, Zeng presented a sequence-to-sequence model, however it was unable to produce multiword entities [25][24]. Nayak and Ng provide an enhancement by using an encoder decoder model that extracts words simultaneously, similar to the machine translation approach. [26][27].

The newer method is table filling based methods[33]. This approach will maintain a table for each relationship, with the entries typically representing the beginning and ending locations of the two entities that have this particular relationship [28]. Consequently, the RTE work is changed into a task that accurately fills these tables [29]. Wang [31] proposed TPLinker based on table filling that considers the problem of mark pair linking in joint extraction and presents a new handshake marking method that aligns the border marks of entities under each connection type [32].

## 3 Method

We proposed a feature reasoning relational triple extraction model (TEFR). The framework of TEFR is shown in Figure.1. The historical features of token pairs are integrated in the framework which can better reasoning implicit features the differences between relationships and token pairs by applying table filling method.

### 3.1 Token Pair Tagging Method

Given a sentence, we maintain a table feature for each relationship. The main component of our approach is to provide each table item a suitable label. Here, the label set is defined as L, including eight kinds of labels. First letter of label represents whether the subject is composed of complex tokens. If complex tokens, it is marked C; If an easy and single token, it is marked E. Similarly, the second letter of label represents whether the object is composed of complex tokens. If it is complex tokens, it is marked as C, and if it is an easy token, it is marked as E. The third letter of label reflects the subject's and object's head and tail positions. If it is the head at the same time, it is marked H; if it is the tail at the same time, it is marked T. In particular, when the subject and object are composed of a token, it is marked as SS, and the rest is marked NULL.

One of the primary benefits of our filling method is that each tag may not only represent the position of a tag in the subject or object, but also represent the type of the entity, such as complex entity and easy entity. As a result of the increased information carried by each tag, the total number of things to be completed is generally modest.

### 3.2 TEFR Method Detail

Figure 1 shows the main architecture of TEFR. It is made up of four major modules: encoder module, table features filling module, feature reasoning module and triple extracting module. Execute table features filling module and feature reasoning module several times through iteration, and gradually refine the table features. Finally, based on the updated table attributes, triple extracting module fills each table and generates total triples from these filled tables.

In encoder module, the encoder is a pretrained Bert base model. The module initially encodes a sentence into a symbolic representation sequence I. Input I into different feed-forward networks (FNN) and create initial subject features and object features through one-way propagation

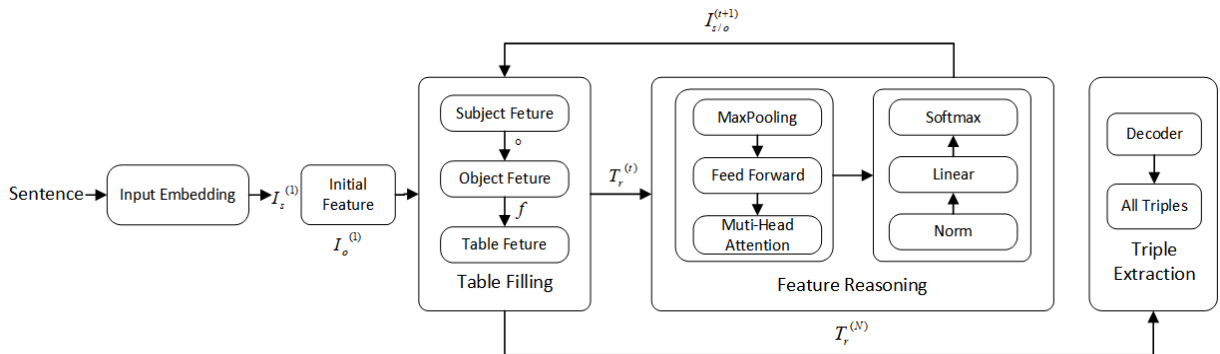

**Figure 1:** Model architecture. $I_{s/o}^{(1)}$ represents initial feature.

(represented by $I_s^{(1)}$ and $I_o^{(1)}$ respectively), as described in Eq (1) and Eq (2).

$$I_s^{(1)} = W_1 I + b_1 \qquad (1)$$

$$I_o^{(1)} = W_2 I + b_2 \qquad (2)$$

In table features filling module, we record the subject and object features at iteration $t$ as $I_s^{(t)}$ and $I_o^{(t)}$. After that, the module accepts them as input and creates a table feature for each item which in predefined set of relationships. The table feature of relationship $r$ at iteration $t$ is marked as $T_r^{(t)}$, and its size is the same as that of table. Each item in $T_r^{(t)}$ represents the tag characteristics of a token pair. Specifically, for a pair $(w_i, w_j)$, we record its tag feature as $T_r^{(t)}(i,j)$, which is calculated by Eq (3).

$$T_r^{(t)}(i,j) = W_r \text{ReLU}(I_{s,i}^{(t)} \circ I_{o,j}^{(t)}) + b_r \qquad (3)$$

where $\circ$ represents the Hadamard Product, $I_{s,i}^{(t)}$ and $I_{o,j}^{(t)}$ represent the subject and object feature of tokens $(w_i, w_j)$ at the t-th iteration respectively.

In feature reasoning module, the module generates implicit subject and object features based on the existing features. The freshly created features will then be returned to TFF for the following iteration. The module specifically comprises the three stages listed below.

In the first stage, table attributes should be combined. Assuming that the round of iteration is t, we integrate all connection table features to create a unified table feature ($T^{(t)}$) which will reveal a lot about token pairs and relationships. We apply the maximum pooling and the FFN model on $T_r^{(t)}$ to create a table feature ($T_s^{(t)}$) and an object related table feature ($T_o^{(t)}$), as shown in Eq (4).

$$T_{s/o}^{(t)} = W_{s/o} \text{maxpool}(T^{(t)}) + b_{s/o} \qquad (4)$$

Where $W_{s/o}$ is the trainable weight and $B_{s/o}$ is the trainable bias. The maximum pool is used to extract important features of topics and objects from existing features.

In the second stage, we mine expected reasoning features. Here, we mainly use the model based on transformer to reason the implicit features between relationships and token pairs. We use the multi-head-self-attention approach on $T_{s/o}^{(t)}$ to mine implicit relationships. The sentence relationship is also used as part of the input. Next, we use FFN to generate new subject and object features. The whole process of subject object feature reasoning can be written with the following Eq (5), (6), (7).

$$T_{s/o}^{(t)} = \text{MultiHeadSelfAtt}(T_{s/o}^{(t)}) \qquad (5)$$

$$\hat{I}_{s/o}^{(t+1)} = \text{MultiHeadAtt}(T_{s/o}^{(t)}, I, I) \qquad (6)$$

$$I_{s/o}^{(t+1)} = \text{ReLU}(\hat{I}_{s/o}^{(t+1)} W + b) \qquad (7)$$

In the third stage, in order to avoid the vanishing gradient problem and develop the final theme and object features, we employ the residual network. Finally, the features of the subjects and objects are sent to table features filling module as the features in the next iteration.

In triple extracting module, taking the last iteration's table feature ($T^{(N)}$) as input, it returns all triples. Then, triple extracting module decodes the filled table to derive all triples. Specifically, for each relationship, first fill its table with the method shown in Eq (8) and Eq (9).

$$\widehat{table}_r(i,j) = \text{soft} max(T_r^{(N)}(i,j)) \qquad (8)$$
$$table_r(i,j) = \underset{l \in L}{argmax}(\widehat{table}_r(i,j)[l]) \qquad (9)$$

The triple extracting module then decodes all triples by decoding the entire table. In our decoding method, different search paths are created according to different token pair types. If the subject and object are complex entities, the search order from beginning to end will be based on the direction of CCH and CCT in the table, and the search order from end to end will be based on the direction of CCT and CCH in the table. If it is a simple entity pair, it will be searched according to the EE tag in the table.

### 3.3 Loss Function

The model loss is defined as follows.

$$L = \sum_{i=1}^{n} \sum_{j=1}^{n} \sum_{r=1}^{R} - \log p(y_{r,(i,j)} = table_r(i,j)) \quad (10)$$

Where $y_r(i,j)$ is the truth index label of $(w_i, w_j)$ for relation $r$.

## 4 Experiments

### 4.1 Dataset

In order to facilitate the comparison of our model with previous work, we followed the popular data set selection: NYT, WebNLG. According to the annotation standard, NYT and WebNLG have two versions: 1) annotate the entity's final word, and 2) annotate the span of the entire entity. The data set of the first version is expressed as NYT* and WebNLG*. In addition to the above four datasets, we also used TFH_ Annotated_ Dataset as technological patents datasets. TFH_ Annotated_Dataset is an annotated patent data set related to thin film head technology in hard disk. which is used to annotate semantic relationships between entities. The well-designed information pattern for patent annotation contains 15 semantic relationships. The basic information of these datasets is shown in Table 1.

**Table 1:** Properties of datasets used for experiment

| Category | NYT24/NYT24* | | WebNLG | | WebNLG* | | TFH_Annotated_Dataset | |
|---|---|---|---|---|---|---|---|---|
| | Train | Test | Train | Test | Train | Test | Train | Test |
| **Normal** | 37013 | 3266 | 1596 | 246 | 1596 | 246 | 3259 | 722 |
| **ALL** | 56195 | 5000 | 5019 | 703 | 5019 | 703 | 3259 | 722 |
| **Relation** | 24 | | 216 | | 171 | | 15 | |

### 4.2 Baselines

For comparison, we use the following model as the benchmark: (1) ETL-Span [7] using span based labeling scheme, the joint extraction task is decomposed into several sequence label problems. This model can fully capture the semantic dependency between different steps, and reduce the noise of unrelated entity pairs; (2) CasRel [33] is a new cascading binary annotation framework model, in which the relationship is modeled as a function that maps from the beginning entity to the end entity, which turns the previous classification task into the problem of finding triples0; (3) Based on the Bert decoder, TPLinker is best performing model on the NYT and WebNLG datasets [31]. It has obvious improvement in dealing with difficult sentences with overlapping relations or sentences with more than two relations. The majority of these baseline experimental results are taken straight from their original journals. In this scenario, we show the best results of the above baseline models. For consistency, we refer to the model proposed by us as TEFR.

### 4.3 Evaluation Metrics

In our experiments, If the extracted triplet's subject entity and object entity's relationship and header are accurate, the extracted triplet is regarded correct. We followed the popular selection reporting standards of micro precision (prec.), recall (rec.), and F1 scores to meet all baselines.

### 4.4 Main Results and Analysis

The primary results are in the Table 2, indicating that TEFR is extremely effective. When compared to the model utilizing the same encoder (BERT based encoder), it obtains the top result on most of datasets. The results also reveal that TEFR performed better on NYT24 and WebNLG, with F1 scores improving by around 0.7% and 3.9%, respectively, as compared to baselines. In contrast, its F1 score was 0.9%, 1.7%, and 4.0% higher than the previous best model on NYT24*, WebNLG* and TFH_Annotated_Dataset.

We can also observe that when compared to the previous best model, the performance improvement of TEFR on WebNLG is more than that on other datasets. For example, we believe this is mostly due to the fact that WebNLG has considerably more relations than NYT. This means that there are more implicit associations to be inferred. Compared with other datasets, TEFR does not perform as well on TFH_Annotated_Dataset. The main reason is that the number of TFH_Annotated_Dataset is small, and manual annotation is a time-consuming and labor-consuming work. The amount of data that can be trained on the model is not as large as other data sets. However, compared with the previous benchmark model, TEFR still performs better.

### 4.5 Detailed Results

We conducted detailed experiments from the following two aspects to prove the effectiveness of our model.

The influence of the number of iterations N was analyzed, and the results are displayed in and Figure 2,3,4, from which the following conclusions may be taken.

On all data sets, when N = 2, feature reasoning module starts to play a role, and the performance of TEFR has been significantly improved, which again shows that the iterative use of historical table features can significantly improve the performance of the model. In the case of N = 3, the performance of TEFR is the best. However, when N > 3, the performance of TEFR decreases, which indicates that the more iterations, inferable information will not increase. This also shows that feature reasoning can only play its best role in a specific number of iterations, and the more iterations, the more useful implicit information cannot be inferred importance of using historical table features to infer hidden features to fill the table. We can further find that in WebNLG and TFH_Annotated_Dataset, the F1 score of was 1.4% and 0.4% higher than that of TPLinker.

**Table 2**: Comparison of RTE performance on datasets

| Model | NYT24* | | | NYT24 | | | WebNLG* | | | WebNLG | | | TFH_Annotated_Dataset | | |
|---|---|---|---|---|---|---|---|---|---|---|---|---|---|---|---|
| | Prec. | Rec. | F1 | Prec. | Rec. | F1 | Prec. | Rec. | F1 | Prec. | Rec. | F1 | Prec. | Rec. | F1 |
| ETL-Span | 84.9 | 72.3 | 78.1 | 85.5 | 71.7 | 78.0 | 84.0 | 91.5 | 87.6 | 84.3 | 82.0 | 83.1 | 74.5 | 57.9 | 65.2 |
| CasRel | 89.7 | 89.5 | 89.6 | 89.8 | 88.2 | 89.0 | **93.4** | 90.1 | 91.8 | 88.3 | 84.6 | 86.4 | 77.0 | 68.0 | 72.2 |
| TPLinker | 91.3 | 92.5 | 91.9 | 91.4 | **92.6** | 92.0 | 91.8 | 92.0 | 91.9 | 88.9 | 84.0 | 86.7 | 78.0 | 68.1 | 72.7 |
| **TEFR** | **92.5** | **93.0** | **92.8** | **92.9** | 92.5 | **92.7** | 93.2 | **93.9** | **93.5** | **92.0** | **88.2** | **90.1** | **78.8** | **72.7** | **75.6** |

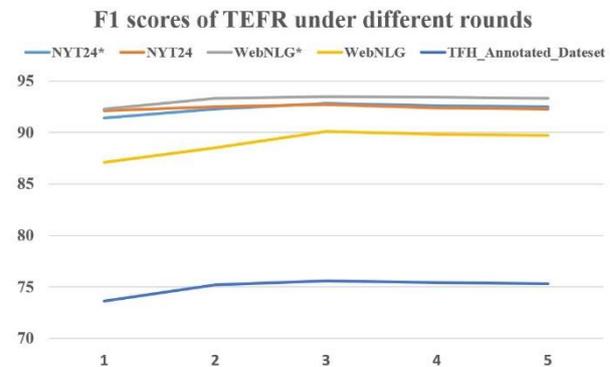

**Figure 2**: F1 scores of TEFR under different rounds

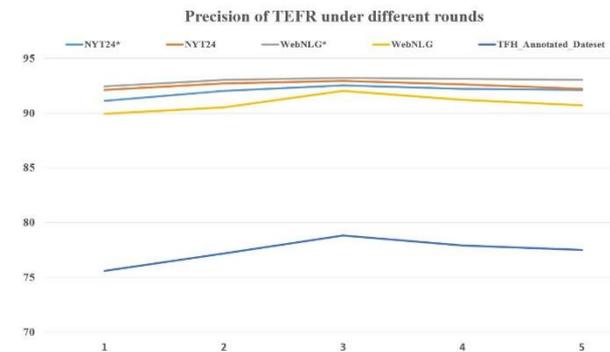

**Figure 3**: Precision of TEFR under different rounds

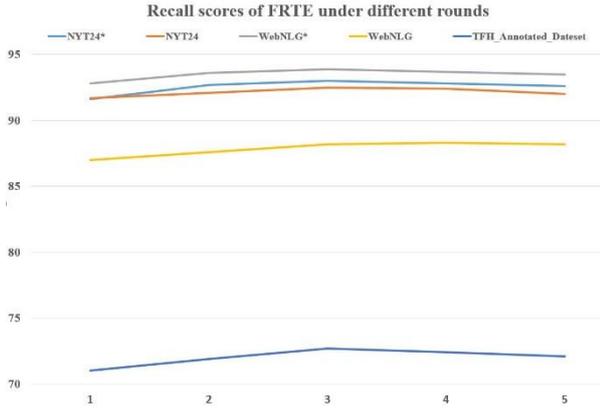

**Figure 4**: Recall of TEFR under different rounds

The result of $FRTE_{w/oMH}$ is lower than that of TEFR, which indicates that the multi-head attention method is effective for reasoning implicit feature. The disadvantage of the single-head attention method is that the model will excessively focus on its own position when encoding the current position information. The multi-head attention method can solve this problem which conducts feature mining from different angles, and the information expressed by the feature is richer than that of the single head self-attention mthod.

**Table 3:** Ablation experiments of TEFR

| Model | NYT24 | | | WebNLG | | | TFH_Annotated_Dataset | | |
|---|---|---|---|---|---|---|---|---|---|
| | Prec. | Rec. | F1 | Prec. | Rec. | F1 | Prec. | Rec. | F1 |
| TPLinker | 91.4 | **92.6** | 92.0 | 88.9 | 84.0 | 86.7 | 78.0 | 68.1 | 72.7 |
| **TEFR** | **92.9** | 92.5 | **92.7** | **92.0** | **88.2** | **90.1** | **78.8** | **72.7** | **75.6** |
| $FRTE_{w/oFR}$ | 91.6 | 92.3 | 92.0 | 89.3 | 86.7 | 88.0 | 78.0 | 68.6 | 73.0 |
| $FRTE_{w/oMH}$ | 91.8 | 92.4 | 92.1 | 90.8 | 87.4 | 89.1 | 78.6 | 71.4 | 74.8 |
| $FRTE_{w/oTF}$ | 92.4 | 92.5 | 92.4 | 91.3 | 87.6 | 89.4 | 78.6 | 72.4 | 75.3 |

### 4.6 Ablation Experiments

We conducted ablation experiments to evaluate the role of some important modules in TEFR. The modules involved are as follows:

The feature reasoning module was completely removed

The result of $FRTE_{w/oTF}$ is higher than that of TPLinker. Both TEFR and TPLinker retrieve triples from table features. The primary distinction between them is the table filling method. As a result, these results indicate the efficacy of table filling method of TEFR.

**Table 4:** F1 scores phrases with varying overlapping patterns and triplet numbers.

| Model | NYT24* | | | | | WebNLG* | | | | |
|---|---|---|---|---|---|---|---|---|---|---|
| | Overlapping Pattern | | | Triplet Number | | Overlapping Pattern | | | Triplet Number | |
| | Normal | SEO | EPO | T=1 | T=2 | Normal | SEO | EPO | T=1 | T=2 |
| CasRel | 87.3 | 91.4 | 92.0 | 88.2 | 90.3 | 89.4 | 92.0 | 94.7 | 89.3 | 90.8 |
| TPLinker | 90.1 | 93.4 | 94.0 | 90.0 | 92.8 | 87.9 | 92.5 | 95.3 | 88.0 | 90.1 |
| SPN | 90.8 | 94.0 | 94.1 | **90.9** | 93.4 | 89.5 | 94.1 | 90.8 | **89.5** | 91.3 |
| **TEFR** | **91.1** | **94.2** | **94.4** | 90.8 | **93.7** | **90.1** | **94.3** | 95.3 | 89.3 | **92.5** |

from the TEFR for assessing the contribution of feature reasoning module. As with previous methods based on table filling, only extracts triples based on the initial table features.

To analyze the contribution of multi-headed attention, replace the multi-headed attention method in feature reasoning module with the single-headed attention method.

Replace the filling policy and decoding policy of TEFR with the filling policy and decoding policy of TPLinker,

Table 3 shows that the performance of $FRTE_{w/oFR}$ is much lower than that of TEFR, which confirms the

### 4.7 Analyses on Overlapping Dataset

We examined TEFR's ability to extract triples from phrases containing overlapping triples and multiple triples. When compared to the previous top models. We categorize phrases based on their degree of overlap and the number of triples they include.

Table 4 displays the results. we can observe that TEFR performed best on all datasets which contains overlapping sentences, and TEFR performed better on practically all sentences including multiple triples. On the NYT24 * and WebNLG * datasets, When, TEFR has a lower F1 score than SPN. One important reason for this is that there are fewer associations between token pairs when the number of triplets is small, which slightly reduces the performance of TEFR.

## 5 Conclusion

We proposed a new RTE method based on feature reasoning in this research. The method we proposed based on the features of existing token pairs and table relationships, reasoning the implicit features of subject and object, and enhances the accuracy of triple extraction.

The method can extract triples from complicated phrases that comprise numerous or overlapping triples as well. The model TEFR is tested on different datasets. A vast number of experiments suggest TEFR is always better than all relatively strong baselines, and has obtained the best results. Furthermore, our model has a fast reasoning speed and a small number of parameters.

## Acknowledgements

This work was supported by the National Natural Science Foundation of China (No.62192784, No.62172056).